\documentclass[a4paper,11pt]{article}
\pdfoutput=1 

\usepackage{jcappub} 

\usepackage[T1]{fontenc} 

\usepackage{natbib}
\title{\boldmath AGN Line-Intensity Mapping: A Probe of Faint Black Holes at Cosmic Noon}

\author[a]{Eli Visbal}
\author[b]{and Greg L. Bryan}

\affiliation[a]{Department of Physics and Astronomy and Ritter Astrophysical Research Center, University of Toledo, 2801 W. Bancroft Street, Toledo, OH, 43606, USA}

\affiliation[b]{Department of Astronomy, Columbia University, 550 West 120th Street, New York, NY, 10027, USA}

\emailAdd{Elijah.Visbal@utoledo.edu}

\abstract{We propose line-intensity mapping (LIM) as a new probe of 
active galactic nuclei (AGN). By cross-correlating 
[Ne\,\textsc{v}] intensity maps with galaxy redshift surveys, we show that 
the cumulative AGN line emission can be detected even when individual 
sources are below the detection threshold. The 97.1\,eV ionization 
potential of [Ne\,\textsc{v}] makes it an essentially uncontaminated tracer 
of AGN activity, arising from the narrow line region which is accessible 
even in heavily obscured AGN. We forecast signal-to-noise ratios using 
a Fisher matrix formalism for two hypothetical future instruments: 
a CDIM-like instrument targeting [Ne\,\textsc{v}]\,$\lambda 3426$ and a 
PRIMA-like instrument optimized for LIM targeting [Ne\,\textsc{v}]\,$14.3\,\mu$m. For the 
CDIM-like case we find strong constraints on the product of the mean 
AGN intensity and bias, $S_{\rm NeV} b_{\rm NeV}$, across $z=2$--$3$, 
with redshift-space distortions enabling individual constraints on 
$S_{\rm NeV}$ and $b_{\rm NeV}$. The LIM signal retains 
sensitivity to AGN below the $5\sigma$ direct detection threshold, 
which at $z=3$ corresponds to $L_{\rm bol} \sim 5\times10^{43}$\,erg\,s$^{-1}$ 
and coincides with the faint end of existing luminosity function 
measurements. Roughly 10\% of the total signal originates from below 
this threshold, with the sub-threshold population detectable at 
$S/N=9~$--$~4$ across $z=2~$--$~3$ (for $S_{\rm NeV} b_{\rm NeV}$). The PRIMA-like instrument achieves 
slightly lower signal-to-noise but provides a complementary probe 
of the AGN population due to the insensitivity of the $14.3\,\mu$m line to 
dust attenuation. AGN LIM can potentially be applied to several scientific problems
including tracing the total AGN emissivity history, constraining the 
black hole-halo connection at faint luminosities, and discriminating 
between supermassive black hole seeding mechanisms.}

\begin{document}
\maketitle
\flushbottom

\section{Introduction}
Line-intensity mapping (LIM) is an emerging observational technique that measures the clustering of spatial fluctuations in line emission within large-scale cosmological volumes \citep{2010JCAP...11..016V, 2026arXiv260203011C}. A strength of LIM is that it probes the cumulative emission from all sources in a volume, 
including those too faint to be detected individually. 
A number of LIM efforts are either planned or underway. This includes observing 21cm radiation with instruments such as \emph{MWA}, \emph{LOFAR}, \emph{HERA} and \emph{SKA} \citep{refId0,DeBoer2017, koopmans, 2022JATIS...8a1007B}.  It also includes LIM of Ly$\alpha$, H$\alpha$, CO, and [C {\sc ii}] with instruments like \emph{SPHEREx} \citep{2014arXiv1412.4872D}, \emph{CDIM} \citep{2019BAAS...51g..23C}, \emph{FYST} \citep{2022A&A...659A..12K}, \emph{COMAP} \citep{2022ApJ...933..182C}, \emph{TIME} \citep{2021ApJ...915...33S}, and \emph{CONCERTO} \citep{2020A&A...642A..60C}.
Up to this point, these efforts have primarily been focused on probing galaxies, reionization, or constraining cosmological parameters. 
In this paper, we propose LIM as a 
probe of active galactic nuclei (AGN). This could open a window to faint black hole populations that are difficult to observe directly.

AGN, powered by accretion onto supermassive black holes, play a 
fundamental role in galaxy evolution through feedback processes that 
regulate star formation \citep[e.g.,][]{1998A&A...331L...1S, 2006MNRAS.365...11C}, however observations of faint AGN remain challenging. For example, at $z=3$, the AGN luminosity function (LF) reported in \cite{2020MNRAS.495.3252S, 2024A&A...685A..97P} are based on data that extend down to $L_{\rm bol} \sim 
3 \times 10^{43}~{\rm erg\,s^{-1}}$. As shown below, there exists significant black hole activity in fainter AGN which can be accessed through LIM.

We focus on the [Ne\,\textsc{v}] lines, arising from Ne$^{4+}$ ions 
requiring photons above 97.1\,eV for their production. This ionization 
potential is sufficiently high that stellar populations contribute 
negligibly; even Wolf-Rayet stars produce few photons above the 
He\,\textsc{ii} edge at 54.4\,eV, and essentially none above 97.1\,eV. This  makes  [Ne\,\textsc{v}] a direct tracer of AGN activity. 
We consider two transitions observable with hypothetical near-future facilities: 
the UV line at 3426\,\AA, accessible to a CDIM-like 
instrument \citep{2019BAAS...51g..23C}, and the mid-infrared 
line at $14.3\,\mu$m, accessible to an instrument similar to PRIMA \cite{2025JATIS..11c1628G}, but optimized for LIM. For each, we explore their AGN LIM capabilities at $z=2-5$ below. We also note that [Ne\,\textsc{v}] emission arises from the narrow line 
region (NLR), which extends beyond the obscuring torus, meaning 
it is accessible even in heavily obscured AGN. While the UV 
$ 3426$\,\AA\ line is subject to some attenuation by dust as 
it escapes the host galaxy, the mid-infrared $14.3\,\mu$m line is 
essentially unaffected by dust extinction, making it a completely 
unobscured probe of AGN activity.

For the instruments we consider, the [Ne\,\textsc{v}] intensity map is faint and contaminated by interloping emission lines at other 
redshifts. We thus consider the cross-correlation between AGN intensity maps and  galaxy redshift surveys. The galaxy field provides a bright signal that improves the signal-to-noise and removes the interlopers in cross-correlation \cite{2010JCAP...11..016V}. 
This allows the faint AGN clustering signal 
to be extracted even when the intensity map itself is noise-dominated. 
We forecast signal-to-noise ratios for both instrument concepts and 
demonstrate that the LIM signal carries sensitivity to AGN below the 
threshold where direct detections can reach. We note that cross-correlation could also be performed with other large-scale structure probes such as galaxy weak lensing or cosmic microwave background (CMB) lensing, but we defer exploring such possibilities to future work.

This paper is organized as follows. In Section~2 we 
describe our methods, including the theoretical framework, signal 
model, and noise model. In Section~3 we present our 
results. In Section~4 we discuss our results and present our 
conclusions. Throughout this work, we assume a $\Lambda {\rm CDM}$ cosmology with parameters consistent with \cite{2020A&A...641A...6P}: $\Omega_{\rm m} = 0.32$, $\Omega_{\Lambda} = 0.68$, $\Omega_{\rm b} = 0.049$, $h=0.67$, $\sigma_8=0.81$, and $n_{\rm s} = 0.96$.

\section{Methods}
\subsection{Power Spectra}

We consider two observables, the cross-power spectrum  
between a [Ne\,\textsc{v}] intensity map and a galaxy redshift survey, $P_{\rm NeV,g}$, and the galaxy auto-power spectrum $P_{\rm g}$. The [Ne\,\textsc{v}] 
auto-power spectrum $P_{\rm NeV}$ is in principle also measurable but 
is contaminated by interloping emission lines from other 
redshifts that are mapped to the same observed frequency (e.g., 
H$\beta$), making it unreliable as a primary 
observable. The cross-power spectrum mitigates this contamination 
since interlopers are uncorrelated with the galaxy field at the target redshift.

The power spectra are given by
\begin{align}
P_{\rm NeV,g} &= S_{\rm NeV}(b_{\rm NeV} + f\mu^2)(b_g + f\mu^2)P_m \\
P_{\rm g} &= (b_g + f\mu^2)^2 P_m + \frac{1}{\bar{n}_g} \\
P_{\rm NeV} &= S_{\rm NeV}^2(b_{\rm NeV} + f\mu^2)^2 P_m + 
P_{\rm NeV}^{\rm shot} + P_{\rm int}
\end{align}
where $P_m$ is the matter power spectrum, $\mu$ is the cosine 
of the angle between the line of sight and the wavevector $\vec{k}$, and 
$f = d\ln D/d\ln a$, where $D$ is the linear 
growth factor and $a$ is the scale factor. We approximate 
$f \approx \Omega_m(z)^{0.55}$ following \cite{2005PhRvD..72d3529L}. Here 
$\bar{n}_g$ is the mean galaxy number density, 
$P_{\rm NeV}^{\rm shot}$ is the AGN shot power, and $P_{\rm int}$ 
is the interloper power (which in principle could come from numerous lines). We have included large-scale redshift-space distortions, which are captured in the factors with $\mu$ dependency \cite{1987MNRAS.227....1K}. The mean intensity $S_{\rm NeV}$ and shot 
noise $P_{\rm NeV}^{\rm shot}$ are computed by integrating over the 
observed AGN luminosity function, as described below. We assume the cross-shot noise signal between the 
AGN and galaxy populations is negligible, this assumption is conservative in the sense that including it would add extra signal in our forecasts below.  We also use the linear power spectrum, $P_{\rm m}$ throughout our calculations. Non-linear effects would primarily boost the 
signal on small scales, making our forecasts conservative. 

The uncertainty on $P_{\rm NeV,g}$ for a single $k$-mode is given by
\begin{equation}
\sigma^2(P_{\rm NeV,g}) = \frac{P_{\rm NeV,g}^2 + 
(P_{\rm NeV} + P_N)P_{\rm g}}{2}
\end{equation}
where $P_N$ is the instrumental noise power \cite{2010JCAP...11..016V}. When computing the error in a $k$-bin, we account for the number of 
available modes $N_k$ via an inverse variance-weighted average,
\begin{equation}
\frac{1}{\sigma^2_{\rm bin}(P_{\rm NeV,g})} = \sum_{\mu} 
\frac{N_k}{\sigma^2(P_{\rm NeV,g})}
\end{equation}
where the sum is over $\mu$-bins within the $k$-bin, and 
\begin{equation}
N_k = \frac{V_{\rm surv}}{(2\pi)^3} 2\pi k^2 \Delta k \sin\theta \, \Delta\theta
\end{equation}
is the number of modes in a shell of width $\Delta k$ at wavenumber 
$k$ and angle $\theta = \cos^{-1}\mu$, with $V_{\rm surv}$ the survey 
volume.

In our forecasts below, we consider two cases. In the first we 
conservatively neglect redshift-space distortions by setting $\mu = 0$, and compute the 
total signal-to-noise on the cross-power,
\begin{equation}
(S/N)^2 = \sum_k \frac{P_{\rm NeV,g}^2}{\sigma^2_{\rm bin }(P_{\rm NeV,g})},
\end{equation}
which corresponds to the constraint on the product 
$S_{\rm NeV} b_{\rm NeV}$. In the second case we include redshift-space distortions, which 
introduce a $\mu$-dependent anisotropy in $P_{\rm NeV,g}$ and 
$P_{\rm g}$ that breaks the degeneracy between $S_{\rm NeV}$ and 
$b_{\rm NeV}$, allowing them to be constrained individually. We use 
the Fisher matrix formalism described in Appendix~\ref{app:fisher} 
to forecast the marginalized constraints on each parameter in this 
case. In both cases, we have assumed that cosmology is known precisely as the astrophysical uncertainties are expected to be much larger.

\subsection{AGN Signal Model and Galaxy Survey}

We adopt the AGN bolometric luminosity function of \cite{2020MNRAS.495.3252S}, 
which takes the form of a double power law, with parameters tabulated 
across a range of redshifts. Specifically we use the ``local polished 
fits'' given in their Table~3 and perform forecasts with the values at $z=2$, 3, 4, and 5.

We assume that the [Ne\,\textsc{v}]\,$\lambda 3426$ luminosity is related 
to the AGN bolometric luminosity via the empirical calibration of 
\cite{2024A&A...685A.141B}, who derive a relation from a sample 
of 94 [Ne\,\textsc{v}]\,$\lambda 3426$-selected type~2 AGN in the COSMOS 
field at $z = 0.6$--$1.2$,
\begin{equation}
\log L_{\rm NeV,3426} \, [\rm{erg\,s}^{-1}] = 0.69 \log L_{\rm bol} 
\, [\rm{erg\,s}^{-1}] + 10.
\end{equation}
For the [Ne\,\textsc{v}]\,$14.3\,\mu$m line we adopt the calibration of 
\cite{2016MNRAS.458.4297G},
\begin{equation}
\log L_{\rm NeV,14} \, [\rm{erg\,s}^{-1}] = 0.66 \log L_{\rm bol} 
\, [\rm{erg\,s}^{-1}] + 11.25.
\end{equation}
Both relations are derived from observed fluxes and therefore already 
account for dust extinction (though this is mainly expected to impact the $\lambda 3426$ line). The mean [Ne\,\textsc{v}] intensity and shot 
power are then
\begin{equation}
S_{\rm NeV} = \int \frac{L_{\rm NeV}(L_{\rm bol})}{4\pi D_L^2} 
\frac{dn}{dL_{\rm bol}} \, y D_A^2 \, dL_{\rm bol},
\end{equation}
and
\begin{equation}
P_{\rm NeV}^{\rm shot} = \int \left(\frac{L_{\rm NeV}(L_{\rm bol})}
{4\pi D_L^2}\right)^2 \frac{dn}{dL_{\rm bol}} 
\left(y D_A^2\right)^2 dL_{\rm bol},
\end{equation}
where $D_L$ and $D_A$ are the luminosity and comoving angular diameter 
distances respectively, and $y = d\chi/d\nu$ is the conversion factor 
between frequency and comoving distance.

The AGN large-scale clustering bias, $b_{\rm NeV}$, is computed via a chain of  relations connecting the bolometric luminosity to the host 
halo mass. We note that this bias model is intended as a rough approximation to provide a reasonable fiducial model.
We convert $L_{\rm bol}$ to black hole mass assuming 
Eddington-limited accretion,
\begin{equation}
M_{\rm BH} = \frac{L_{\rm bol}}{1.2\times10^{38}\,\lambda_{\rm Edd}} 
\, M_\odot
\end{equation}
where we take $\lambda_{\rm Edd} = 1$. The stellar mass of the host 
galaxy is then estimated via $M_* = 1000\,M_{\rm BH}$ \citep{KormendyHo2013}, and the halo 
mass is determined by assuming a star formation efficiency of 10\%, 
such that $M_{\rm halo} = M_* / (0.1 \times \Omega_b/\Omega_m)$ \citep{Behroozi2013}. The 
large-scale bias is computed from the halo mass using the fitting 
formula of \cite{2001MNRAS.323....1S}, and the luminosity-weighted mean bias is given by
\begin{equation}
\label{eqn:bias}
b_{\rm NeV} = \frac{\int b(L_{\rm bol})\, L_{\rm NeV}(L_{\rm bol})\, 
(dn/dL_{\rm bol})\, dL_{\rm bol}}{\int L_{\rm NeV}(L_{\rm bol})\, 
(dn/dL_{\rm bol})\, dL_{\rm bol}}.
\end{equation}
We note that the exact accretion rate assumed does not strongly impact our results. Reducing the assumed Eddington fraction to $\lambda_{\rm Edd} = 0.1$ only improves our signal-to-noise at the percent level. This is because the halos which contribute to the signal are at masses where the clustering bias does not change rapidly as a function of halo mass.

As described above, we consider AGN intensity maps being cross-correlated with galaxy surveys to remove interlopers. For our forecasts presented below, we assume a galaxy survey measured directly with the CDIM-like instrument. The number density of galaxies is computed with the halo mass function from \cite{2001MNRAS.323....1S}. For simplicity, we assume that the star formation rate of galaxies is proportional to the halo mass with a normalization that satisfies the fit to the Madau plot~\cite{Madau2014} with
\begin{equation}
    {\rm SFRD} =0.015\frac{(1+z)^{2.7}}{1+\left[(1+z)/2.9\right]^{5.6}}\mathrm{~M_\odot yr^{-1}Mpc^{-3}}.
\end{equation}
We include all galaxies that are detected at 5$\sigma$ confidence in the H$\alpha$ line (the sensitivity of our assumed instrument is described in the next subsection). We assume the H$\alpha$ luminosity is given by \cite{1998ARA&A..36..189K} $L_{\textrm{gal,H}\alpha}=1.27\times10^{41}(1-f_{\rm dust}) \frac{\dot{M_*}}{\text{M}_{\odot}\text{yr}^{-1}}\text{erg s}^{-1}$, where $\dot{M_*}$ is the star formation rate and $f_{\rm dust}=0.6$ is a factor accounting for dust attenuation in the galaxies \cite{2017ApJ...835..273G}. We compute $b_{\rm g}$ similarly to $b_{\rm NeV}$, but without the luminosity weighting and the halo mass function instead of the AGN luminosity function (see Eq.~\ref{eqn:bias}). 

\subsection{Noise Model}
We consider a CDIM-like instrument with a 1.2\,m primary mirror, 
angular resolution of 2\,arcsec, spectral resolution $R = 300$, and 
an instantaneous field of view of 8\,deg$^2$. This is broadly 
consistent with the CDIM mission concept described in 
\cite{2019BAAS...51g..23C}. We assume a survey consisting of $N_{\rm field} 
= 10$ independent fields each observed for a total integration time 
of $t = 10^8$\,s, giving a total survey area of $\sim 70$\,deg$^2$ (though we note our results are not highly sensitive to the number of fields the integration time is split across).
The instrumental noise power spectrum is computed assuming photon-noise 
limited performance following Eq.~16 in \cite{2016MNRAS.463.3193C}.

We consider three primary interlopers for the [Ne\,\textsc{v}]\,$\lambda 3426$ line: H$\alpha$,
H$\beta$, and [O {\sc iii}] $\lambda5007$. We model H$\beta$ line emission in the same manner as H$\alpha$ for the galaxy survey described above but with the line luminosity reduced by a factor of 0.35 and the $f_{\rm dust}$ changed to 0.72. 
For [O {\sc iii}] we assume $L_{\textrm{gal,OIII}}=1.32\times10^{41}(1-f_{\rm dust}) \frac{\dot{M_*}}{\text{M}_{\odot}\text{yr}^{-1}}\text{erg s}^{-1}$, with $f_{\rm dust} = 0.7$ (these luminosities follow previous LIM calibrations \cite{2017ApJ...835..273G}).

We assume that sources which are individually detected above $5\sigma$ in interloper lines are masked 
from the intensity map prior to computing the interloper power spectrum, removing some contamination before cross-correlation. The residual interloper power from unmasked sources is computed for each line following the geometric remapping 
formalism of \cite{2010JCAP...11..016V}, in which the interloper power spectrum 
is distorted due to the incorrect assumption of the 
target redshift. The remapped wavenumber for each line is given by
\begin{equation}
k' = k \int_0^{\pi/2} \sqrt{\frac{\sin^2\theta}{c_x^2} + \frac{\cos^2\theta}{c_z^2}} \sin\theta \, d\theta
\end{equation}
where $c_x = D_A(z_{\rm int})/D_A(z_{\rm target})$ and 
$c_z = y(z_{\rm int})/y(z_{\rm target})$ are the ratios of the 
transverse and radial distance conversion factors respectively, and 
we have performed a spherical average over angle for simplicity.

The residual interloper power for each line is then
\begin{equation}
P_{\rm int}(k) = \frac{S_{\rm int}^2 b_{\rm int}^2 P_m(k') +P_{\rm shot, int}}{c_x^2c_z},
\end{equation}
where $S_{\rm int}$ is the mean intensity of the unmasked interlopers in a particular line, $b_{\rm int}$ is their luminosity-weighted clustering bias, and $P_{\rm shot, int}$ is their shot power. We leave a full treatment of any additional interlopers for future work, but note that our approach could be conservative if the interlopers could be detected and removed using even brighter lines from the same galaxies that do not contaminate the map as interlopers.

\section{Results}
\label{sec:results}
\begin{figure}
\centering 
\includegraphics[width=12 cm]{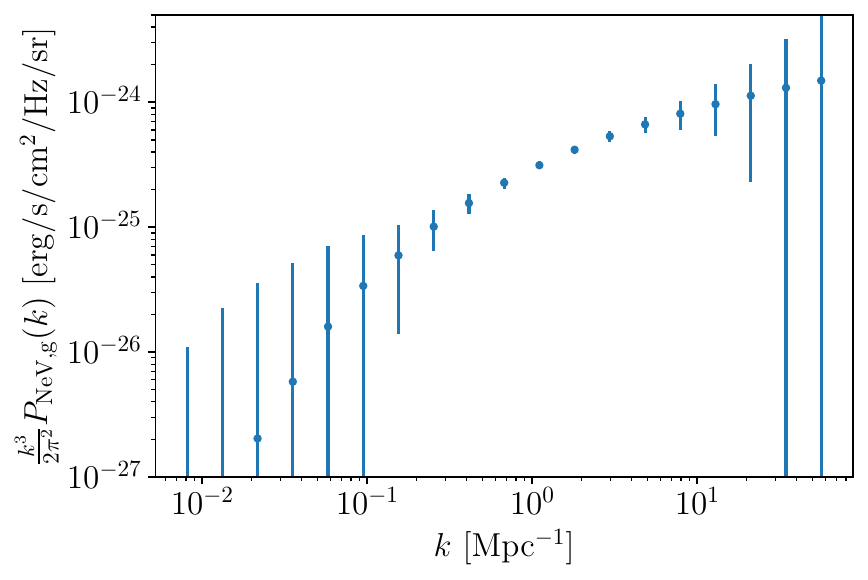}
\caption{\label{fig:cross_power} Cross-power spectrum at $z=3$ for a CDIM-like survey targeting 
[Ne {\sc v}]\,$\lambda 3426$. Error bars show the $1 \sigma$ 
uncertainties and data points denote central values of the assumed k-bins. Note that we conservatively do not include redshift-space distortions here, but that they are considered in the Fisher matrix calculation described in the main text.}
\end{figure}

\begin{figure}
\centering 
\includegraphics[width=12 cm]{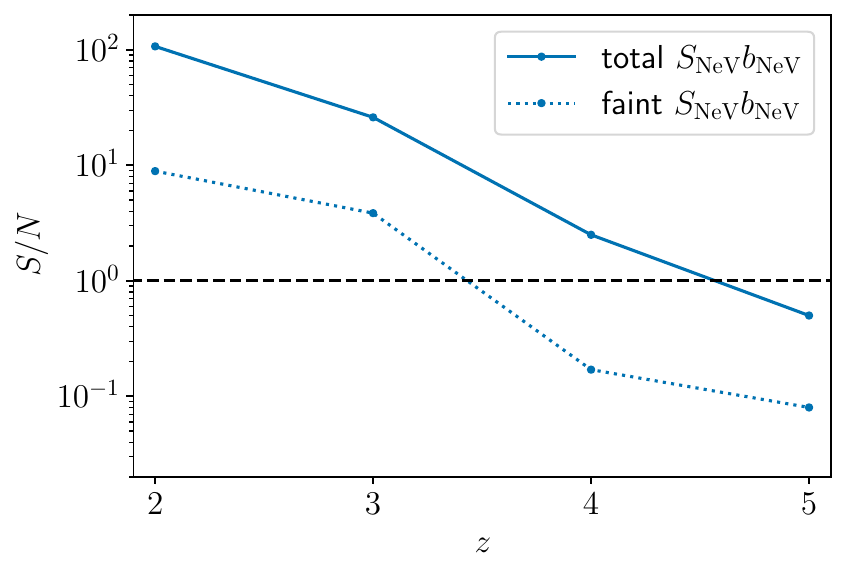}
\caption{\label{fig:sn} Signal-to-noise on $S_{\rm NeV} b_{\rm NeV}$ as a function 
of redshift for a CDIM-like survey cross-correlating galaxies and  [Ne {\sc v}] 
$\lambda 3426$. Both the total AGN population (solid line) and the  population below the CDIM 5$\sigma$ direct detection limit 
(dotted line) are shown.}
\end{figure}

We begin by forecasting results for AGN LIM with the [Ne\,\textsc{v}]\,$\lambda 3426$ line observed with our CDIM-like instrument where we have conservatively ignored redshift-space distortions by setting $\mu=0$. In Figure~\ref{fig:cross_power} we present the cross-power spectrum  $P_{\rm NeV,g}$ as a function of wavenumber $k$ at $z=3$ along with the associated $1\sigma$ error bars. We predict high signal-to-noise across a wide range of scales. 
We show in Figure~\ref{fig:sn} the signal-to-noise on 
$S_{\rm NeV} b_{\rm NeV}$ as a function of redshift for both the 
total signal and the contribution from AGN below the 5$\sigma$ CDIM 
detection threshold.  The total S/N is strong across $z=2$--$3$ and 
falls toward higher redshift, driven by a combination of the declining 
AGN luminosity function and increased cosmological luminosity distance. For the faint-only contribution, the S/N is 9 at $z=2$, 
4 at $z=3$, and undetectable by $z=4$. 

We note that at $z=3$ the CDIM detection threshold for [Ne\,\textsc{v}]\,$\lambda 3426$ corresponds to an AGN luminosity of $L_{\rm bol} \sim 5\times10^{43}$\,erg\,s$^{-1}$. 
This coincides both with the faint end of 
the \cite{2020MNRAS.495.3252S} observational data and thus would probe sources which are currently too faint to study. At $z=2$ the situation is  somewhat different. CDIM can detect sources only down to 
$L_{\rm bol} \sim 1.5\times10^{43}$\,erg\,s$^{-1}$, while 
\cite{2020MNRAS.495.3252S} have direct observational support to even lower 
luminosities, so the LIM measurement at $z=2$ is probing a population 
that is faint but has contribution from detectable sources.

\begin{figure}
\centering 
\includegraphics[width=12 cm]{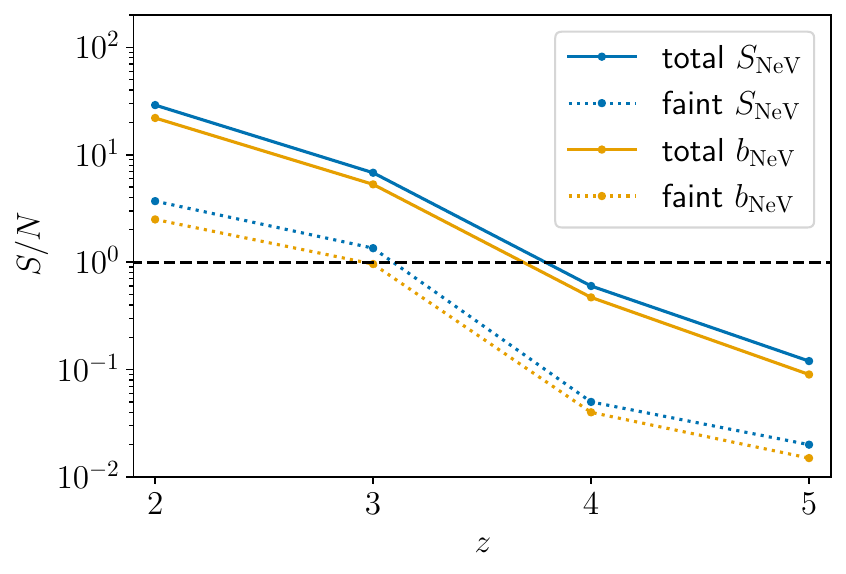}
\caption{\label{fig:sn_rsd} Signal-to-noise on $S_{\rm NeV}$ and 
$b_{\rm NeV}$ using redshift-space distortions as described in Appendix A, for a CDIM-like survey 
cross-correlating galaxies and [Ne {\sc v}]\,$\lambda 3426$. Both total (solid lines) 
and sub-5$\sigma$ detection threshold contributions (dotted lines) are shown. }
\end{figure}

Next, we forecast how well redshift-space distortions can be utilized to break the degeneracy between the mean LIM signal and the clustering bias (see the formalism described in Appendix A). The signal-to-noise on $S_{\rm NeV}$ and $b_{\rm NeV}$  
is shown in Figure~\ref{fig:sn_rsd} as a function of redshift, for 
both the total and faint-only contributions. The signal-to-noise 
is sufficient for meaningful individual constraints at $z=2$--$3$, but not at higher redshifts. 

\begin{figure}
\centering 
\includegraphics[width=12 cm]{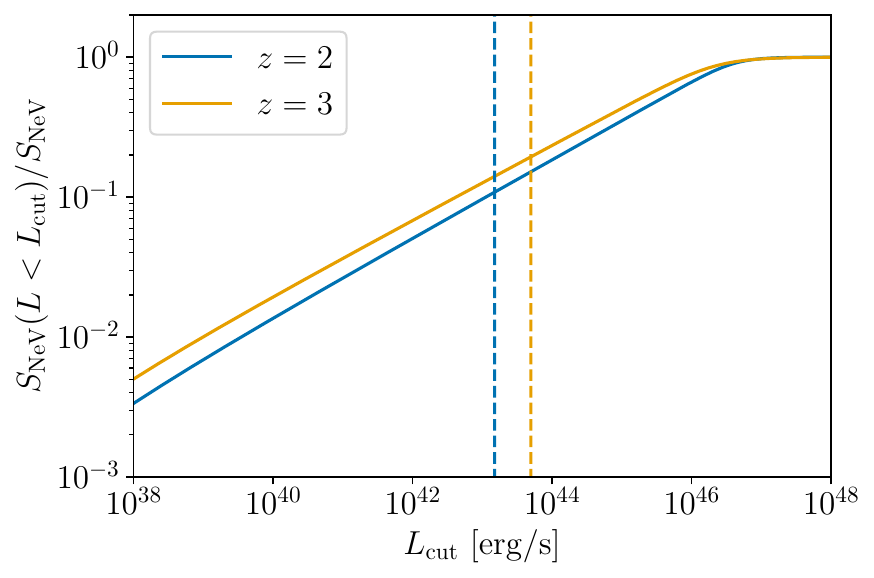}
\caption{\label{fig:cumulative_signal} Cumulative fraction of the total [Ne {\sc v}] mean 
intensity $S_{\rm NeV}$ contributed by AGN below a given bolometric 
luminosity, $L_{\rm cut}$, for a CDIM-like survey targeting [Ne {\sc v}]\,
$\lambda 3426$. The 5$\sigma$ CDIM direct detection thresholds at $z=2$ and  $z=3$ are marked with vertical lines, both of which correspond to 
being above approximately 10\% of the cumulative signal.}
\end{figure}

In Figure~\ref{fig:cumulative_signal} we show the cumulative 
fraction of the total [Ne\,\textsc{v}] mean intensity $S_{\rm NeV}$ 
contributed by AGN below a given bolometric luminosity. The CDIM 
direct detection thresholds at $z=2$ and $z=3$ are marked, 
corresponding to $L_{\rm bol} \sim 1.5\times10^{43}$ and 
$5\times10^{43}$\,erg\,s$^{-1}$ respectively. In both cases, roughly 10\% of the total signal comes from below the detection threshold. The exact fraction depends on the uncertain faint-end slope of the AGN luminosity function, and a measurement 
of $S_{\rm NeV}$ would therefore directly constrain the faint-end 
AGN emissivity in a way that is complementary to and independent of 
direct detection approaches. This shows that LIM is likely to inform us about the aggregate properties of faint AGN near cosmic noon.

We also forecast constraints for a hypothetical PRIMA-like instrument 
optimized for LIM targeting the [Ne\,\textsc{v}]\,$14.3\,\mu$m line, 
which provides a complementary extinction-free probe of the AGN 
population. We assume a 1.8\,m primary mirror cooled to cryogenic 
temperatures with background-limited detectors, a spectral resolution 
of $R\sim 100$, and a survey consisting of $N_{\rm field}=10$ 
independent fields with total integration time $t=10^8$\,s distributed 
across $\sim 1000$ spatial pixels. At $z=3$ we forecast $S/N \sim 6$ 
on $S_{\rm NeV} b_{\rm NeV}$ after mitigation of the 
[Ne {\sc iii}]\,$15.56\,\mu$m interloper via masking of independently 
detected galaxies. For the faint-only contribution below 
$L_{\rm bol} \sim 10^{44}$\,erg\,s$^{-1}$ we forecast $S/N \sim 3$. 
A key advantage of the $14.3\,\mu$m line over [Ne\,\textsc{v}]\,
$\lambda 3426$ is its insensitivity to dust extinction, providing an 
unobscured view of the AGN population that is complementary to the 
CDIM-like measurement. However, achieving high signal-to-noise is 
more challenging with our assumed instrument configuration, and the 
constraints are somewhat weaker than those forecasted for the 
CDIM-like case.
 
\section{Discussion and Conclusions}
\label{sec:discussion}

We have proposed a line-intensity mapping technique that exploits the AGN-exclusive [Ne {\sc v}] emission line to probe the faint black hole population near cosmic noon. By cross-correlating [Ne\,\textsc{v}] intensity maps 
with galaxy redshift surveys, we forecast constraints on 
the mean AGN line emissivity $S_{\rm NeV}$ and large-scale bias 
$b_{\rm NeV}$ for both a CDIM-like instrument targeting 
[Ne\,\textsc{v}]\,$\lambda 3426$ and a hypothetical PRIMA-like instrument 
targeting [Ne\,\textsc{v}]\,$14.3\,\mu$m. We also briefly considered a \emph{SPHEREx}-like instrument 
\citep{2014arXiv1412.4872D} targeting [Ne\,\textsc{v}]\,$\lambda 3426$, 
finding $S/N \sim 1$ at $z=2$ even with an optimistic galaxy survey. This is primarily due to its lower sensitivity, underscoring the need for a dedicated deeper survey for [Ne\,\textsc{v}] AGN LIM to be viable.

For the CDIM-like case, the total S/N on $S_{\rm NeV} b_{\rm NeV}$ 
is strong across $z=2$--$3$ and declines at higher redshift. 
Redshift-space distortions enable individual constraints on 
$S_{\rm NeV}$ and $b_{\rm NeV}$ at $z=2$--$3$, with the degeneracy 
becoming increasingly difficult to break at higher redshift. The 
key science result is the sensitivity of the LIM signal to AGN below 
the $5\sigma$ CDIM direct detection threshold. At $z=3$ this 
threshold corresponds to $L_{\rm bol} \sim 5\times10^{43}$\,erg\,s$^{-1}$, 
coinciding with the faint end of the \cite{2020MNRAS.495.3252S} 
observational data, and roughly 10\% of the total signal originates 
from below this limit. The sub-threshold population is detectable 
at S/N of 9 and 4 at $z=2$ and $z=3$ respectively, and becomes 
undetectable by $z=4$. Measuring this quantity would provide an unbiased census of sources  too faint to appear in any existing survey, free from the selection biases that affect stacking analyses. For the PRIMA-like case we  forecast $S/N \sim 6$ on $S_{\rm NeV} b_{\rm NeV}$ and $S/N \sim 3$ on the faint population below $L_{\rm bol} \sim 10^{44}$\,erg\,s$^{-1}$ at $z=3$. While the constraints are weaker than the CDIM-like case, the $14.3\,\mu$m line would provide a dust extinction-free view of the AGN population that is complementary to the UV line measurement. In particular, the ratio of $S_{\rm NeV}$ measured in the two lines 
provides a direct probe of the mean dust obscuration in the host 
galaxy interstellar medium, averaged over the full population including 
faint and obscured sources that are inaccessible to direct observation.

The choice of [Ne\,\textsc{v}] as the target line is motivated by its 
97.1\,eV ionization potential, which makes it an essentially 
uncontaminated tracer of AGN activity. Furthermore, [Ne\,\textsc{v}] 
emission arises from the narrow line region (NLR), which extends 
beyond the obscuring torus, making it accessible even in heavily 
obscured AGN. He\,\textsc{ii}\,$\lambda 1640$ 
is in principle accessible to CDIM-like instruments, however the 54.4\,eV 
ionization potential is sufficiently low that star-forming galaxies 
may contribute to the mean intensity complicating the interpretation.
Nevertheless, a joint measurement of He\,\textsc{ii}\,$\lambda 1640$ 
and [Ne\,\textsc{v}]\,$\lambda 3426$ would be valuable. 
Since [Ne\,\textsc{v}] traces only AGN while He\,\textsc{ii} receives 
contributions from both AGN and star formation, the combination probes the relative 
contributions of the two populations. At higher ionization potentials, coronal lines are even more 
AGN-exclusive but are typically expected to be fainter. Other potentially interesting lines include C\,\textsc{iv}\,$\lambda 1549$, 
Mg\,\textsc{ii}\,$\lambda 2798$, [O {\sc iv}]\,$25.89\,\mu$m, and 
[Ne {\sc vi}] in the mid-infrared, which we leave for future investigation.

The detectability of AGN LIM opens up several scientific 
applications. By measuring $S_{\rm NeV}$ across multiple redshift 
bins one can trace the total AGN emissivity history including faint 
and obscured sources, providing the AGN analog of the 
star formation rate history \citep{Madau2014}. Since [Ne\,\textsc{v}] 
traces the hard-ionizing flux above 97.1\,eV, $S_{\rm NeV}$ provides 
a direct measurement of the evolution of the hard ionizing background 
near $z\sim3$, when He\,\textsc{ii} reionization is thought to occur, 
without requiring completeness corrections for faint or obscured sources. 
The bias measurement $b_{\rm NeV}$ extends the black hole-halo 
connection into the low luminosity regime. The combination 
of $S_{\rm NeV}$ and $b_{\rm NeV}$ may further constrain the AGN duty 
cycle and triggering mechanisms in faint sources. A joint LIM 
measurement of [Ne\,\textsc{v}]\,$\lambda 3426$ and He\,\textsc{ii}\,$\lambda 1640$ 
would also help disentangle the three contributions to the 
He\,\textsc{ii} intensity mapping signal from AGN, star-forming galaxies, and 
Population~III stars, with the [Ne\,\textsc{v}] measurement providing 
an independent constraint on the AGN contribution that can be 
subtracted to isolate the remaining components. Finally, the faint-end 
AGN emissivity may help discriminate between supermassive black hole 
seeding mechanisms. For example, light seed scenarios arising from 
Population~III stellar remnants may predict a more numerous population 
of faint AGN compared to massive seed scenarios even at $z{\sim}3$, 
leading to different predictions for $S_{\rm NeV}$.

We have demonstrated that AGN line LIM is a viable new 
technique for probing the faint and obscured black hole population 
at cosmic noon. The [Ne\,\textsc{v}] lines provide an essentially 
uncontaminated tracer of AGN activity accessible to near-future 
facilities, with the cross-correlation technique naturally suppressing 
interloper contamination. As next-generation instruments come online, AGN LIM may become an 
important new tool for understanding the demographics of the faint 
and obscured AGN population across cosmic time.

\acknowledgments
We thank Anne Medling, Joaquin Vieira, and Yue Shen for helpful discussions. EV acknowledges the support of NSF grant AST-2009309, NASA ATP grant 80NSSC22K0629, and STScI grant JWST-AR-05238. 
GLB acknowledges support from the NSF (AST-2307419), NASA TCAN award 80NSSC21K1053, and the Simons Foundation through the Learning the Universe Collaboration.
The authors acknowledge the use of Claude (Anthropic) for assistance in editing portions of this manuscript. 

\appendix
\section{Fisher Matrix Formalism}
\label{app:fisher}
We use a Fisher matrix approach to forecast constraints on the mean 
[Ne\,\textsc{v}] intensity $S_{\rm NeV}$ and AGN bias $b_{\rm NeV}$, 
with the degeneracy between these parameters broken by the anisotropic 
signal introduced by redshift-space distortions. For compactness we 
define
\begin{equation}
\beta_{\rm NeV} \equiv b_{\rm NeV} + f\mu^2, \quad 
\beta_g \equiv b_g + f\mu^2,
\end{equation}
so that the power spectra may be written as
\begin{equation}
P_{\rm NeV,g} = S_{\rm NeV}\beta_{\rm NeV}\beta_g P_m, \quad
P_{\rm g} = \beta_g^2 P_m + \frac{1}{\bar{n}_g}, \quad
P_{\rm NeV} = S_{\rm NeV}^2\beta_{\rm NeV}^2 P_m + 
P_{\rm NeV}^{\rm shot} + P_{\rm int}.
\end{equation}
The Fisher matrix is given by
\begin{equation}
F_{ij} = \sum_{k,\mu} \left[ \frac{1}{2} \mathrm{Tr}\left(
\mathbf{C}^{-1} \frac{\partial \mathbf{C}}{\partial \theta_i} 
\mathbf{C}^{-1} \frac{\partial \mathbf{C}}{\partial \theta_j}\right) 
+ \frac{\partial \boldsymbol{\mu}^T}{\partial \theta_i} 
\mathbf{C}^{-1} \frac{\partial \boldsymbol{\mu}}{\partial \theta_j} 
\right],
\end{equation}
where the sum is over $k$ and $\mu$ bins, 
$\boldsymbol{\theta} = \{S_{\rm NeV}, b_{\rm NeV}, b_g\}$ is the 
parameter vector, and the data vector is 
$\boldsymbol{\mu} = (P_{\rm NeV,g}, P_{\rm g})^T$. The covariance 
matrix is
\begin{equation}
\mathbf{C} = \frac{1}{N_k}\begin{pmatrix} 
\frac{1}{2}(P_{\rm NeV,g}^2 + P_{\rm NeV}P_{\rm g}) , & 
P_{\rm NeV,g}P_{\rm g} \\ 
P_{\rm NeV,g}P_{\rm g} , & P_{\rm g}^2 
\end{pmatrix},
\end{equation}
where $N_k$ is the number of modes in each bin. The 
derivatives of the mean vector with respect to each parameter are
\begin{align}
\frac{\partial \boldsymbol{\mu}}{\partial S_{\rm NeV}} &= 
\begin{pmatrix}\beta_{\rm NeV}\beta_g P_m \\ 
0\end{pmatrix}, \quad
\frac{\partial \boldsymbol{\mu}}{\partial b_{\rm NeV}} = 
\begin{pmatrix}S_{\rm NeV}\beta_g P_m \\ 
0\end{pmatrix}, \quad
\frac{\partial \boldsymbol{\mu}}{\partial b_g} = 
\begin{pmatrix}S_{\rm NeV}\beta_{\rm NeV} P_m \\ 
2\beta_g P_m\end{pmatrix}.
\end{align}
The derivatives of the covariance matrix with respect to each 
parameter are
\begin{equation}
\frac{\partial \mathbf{C}}{\partial S_{\rm NeV}} = \frac{1}{N_k}
\begin{pmatrix} 
P_{\rm NeV,g}\beta_{\rm NeV}\beta_g P_m \;+\; 
S_{\rm NeV}\beta_{\rm NeV}^2 P_m P_{\rm g} , & 
P_{\rm g}\beta_{\rm NeV}\beta_g P_m \\ 
P_{\rm g}\beta_{\rm NeV}\beta_g P_m , & 0 
\end{pmatrix},
\end{equation}
\begin{equation}
\frac{\partial \mathbf{C}}{\partial b_{\rm NeV}} = \frac{1}{N_k}
\begin{pmatrix} 
P_{\rm NeV,g}S_{\rm NeV}\beta_g P_m \;+\; 
S_{\rm NeV}^2\beta_{\rm NeV} P_m P_{\rm g} , & 
P_{\rm g}S_{\rm NeV}\beta_g P_m \\ 
P_{\rm g}S_{\rm NeV}\beta_g P_m , & 0 
\end{pmatrix},
\end{equation}
\begin{equation}
\frac{\partial \mathbf{C}}{\partial b_g} = \frac{1}{N_k}
\begin{pmatrix} 
P_{\rm NeV,g}S_{\rm NeV}\beta_{\rm NeV} P_m \;+\; 
P_{\rm NeV}\beta_g P_m , & 
S_{\rm NeV}\beta_{\rm NeV} P_m P_{\rm g} \;+\; 
2P_{\rm NeV,g}\beta_g P_m \\ 
S_{\rm NeV}\beta_{\rm NeV} P_m P_{\rm g} \;+\; 
2P_{\rm NeV,g}\beta_g P_m , & 
4P_{\rm g}\beta_g P_m 
\end{pmatrix}.
\end{equation}
Marginalized constraints on individual parameters are obtained from 
the diagonal of the inverse Fisher matrix, 
$\sigma(\theta_i) = (F^{-1})_{ii}^{1/2}$.

\bibliography{neon_LIM}
\end{document}